\documentclass[aps,prd,reprint,twocolumn,nofootinbib]{revtex4-1}
\pdfoutput=1 
\usepackage{graphicx}
\usepackage{dcolumn}
\usepackage{bm}
\usepackage{color}
\usepackage{textcomp}
\usepackage{subfigure}
\usepackage{feynmp}
\usepackage{amsmath,amssymb,array}
\usepackage{enumerate}
\usepackage{hhline}
\usepackage{hyperref}
\hypersetup{colorlinks=true,urlcolor=[rgb]{0,0,0.5},citecolor=[rgb]{0,0,0.5},linkcolor=[rgb]{0,0,0.5}}

\newcommand{\beqa}{\begin{eqnarray}}
\newcommand{\eeqa}{\end{eqnarray}}
\newcommand{\be}{\begin{equation}}
\newcommand{\ee}{\end{equation}}
\newcommand{\ba}{\begin{array}} 
\newcommand{\ea}{\end{array}}

\begin{document} 
\title{Minimal spontaneous CP-violating GUT and predictions for leptonic CP phases}
\author{Ketan M. Patel}
\email{kmpatel@prl.res.in}
\affiliation{Theoretical Physics Division, Physical Research Laboratory, Navarangpura, Ahmedabad-380009, India.}

\begin{abstract}
A non-supersymmetric renormalizable $SO(10)$ model, with CP invariant Yukawa sector consisting of Lorentz scalars in $10$ and $\overline{126}$ dimensional representations, is proposed. The elemental Yukawa couplings are real due to CP symmetry. The latter is broken in the low energy effective theory through the standard model Higgs which is a complex linear combination of electroweak doublets residing in $10$ and $\overline{126}$ scalars. As a result, the mass matrices in the quark and lepton sectors, including those of heavy and light neutrinos, depend only on three phases which in turn determine CP violation in both sectors. The model is comprehensively analysed for its viability and predictions including the possibility to generate baryon asymmetry through thermal leptogenesis. It predicts relatively small values for CP phases in the lepton sector. Successful leptogenesis further restricts the ranges to $-0.4 \le \sin\delta \le 0.4$  for the Dirac phase and $-0.3 \le \sin \eta_1 \le 0.2$, $-0.5 \le \sin \eta_2 \le 0.5$ for the Majorana phases.
\end{abstract}

\maketitle

\section{Introduction}
\label{sec:intro}
Grand Unified Theories (GUT), through the virtue of quark-lepton unification, often provide a predictive and verifiable theory of flavour by establishing correlations between the masses and mixing pattern of the Standard Model (SM) fermions. This, in particular, has been demonstrated for the renormalizable versions of the $SO(10)$ GUT \cite{Fritzsch:1974nn,GellMann:1980vs} in which the effective mass matrices of the quarks and leptons including neutrinos stem from a few fundamental Yukawa couplings \cite{Aulakh:1982sw,Clark:1982ai,Babu:1992ia,Matsuda:1999yx,Matsuda:2000zp,Aulakh:2003kg,Goh:2003sy,Dutta:2004hp}. Quantitative fits to the known fermion mass spectrum have been carried out for several variants of the basic framework in \cite{Aulakh:2005mw,Babu:2005ia,Bertolini:2005qb,Bertolini:2006pe,Grimus:2006bb,Grimus:2006rk,Bajc:2008dc,Aulakh:2008sn,Aulakh:2007ir,Joshipura:2009tg,Joshipura:2011rr,Joshipura:2011nn,Altarelli:2013aqa,Dueck:2013gca,Meloni:2014rga,Feruglio:2014jla,Feruglio:2015iua,Meloni:2016rnt,Babu:2016bmy,Babu:2016cri,Buchmuller:2017vut,Ohlsson:2018qpt,Boucenna:2018wjc,Babu:2018tfi,Ohlsson:2019sja,Mummidi:2021anm,Fu:2022lrn} which then allow one to derive predictions for the experimentally unknown observables making the underlying framework testable. As an example, a range for the reactor mixing angle, $ 0.015 < \sin^2\theta_{13} < 0.03$, was predicted in a minimal non-supersymmetric $SO(10)$ model in \cite{Joshipura:2011nn} which was found consistent with the value, $ \sin^2 \theta_{13} = 0.023 \pm 0.0023$, directly measured by Daya Bay \cite{DayaBay:2012fng} and RENO \cite{RENO:2012mkc} just a year later.

Having measured all three leptonic mixing angles with reasonably good precision, the efforts in the neutrino experimental activities are now focused to measure the CP violation and the absolute scale of neutrino masses. Therefore, it is worthwhile to derive clear and comprehensive predictions for these observables in realistic and predictive frameworks. Notably, some of the simplest models which have been investigated comprehensively for their predictions for leptonic CP violation predict no clear preference for specific values for the leptonic CP phases, see for example  \cite{Joshipura:2011nn,Feruglio:2014jla,Babu:2018tfi,Mummidi:2021anm}. This can be attributed mainly to the presence of the complex Yukawa couplings whose phases bring enough freedom to accommodate the desired amount of CP violation.

A way to reduce the number of phases in the effective couplings is to impose a suitably defined CP symmetry on the Yukawa sector \cite{Grimus:1995zi}. The symmetry can be broken spontaneously to account for the non-zero CP violation in the quark sector. $SO(10)$ models based on this principle have been constructed and studied earlier in \cite{Grimus:2006bb,Grimus:2006rk,Joshipura:2009tg,Joshipura:2011nn,Fu:2022lrn}. All these models use the most general Yukawa sector consisting of scalars in $10$, $\overline{126}$ and $120$ dimensional irreducible representations of $SO(10)$.  Reduction in parameters is then achieved by imposing additional symmetries and/or assuming specific choices for the phases of the vacuum expectation values (VEV) of the underlying scalars.

In this article, we propose an alternative model based on the CP invariant Yukawa interaction that uses only $10$ and $\overline{126}$ and no additional symmetries and/or ad-hoc assumptions about VEV. The theory below the GUT scale contains an electroweak doublet scalar, to be identified with the SM Higgs, which is a combination of the similar scalars residing in $10$ and $\overline{126}$. Spontaneously broken CP makes this combination complex and introduces three phases in the effective Yukawa couplings of the quarks and leptons. These phases determine CP violation in both the quark and lepton sectors. Exploiting the predictive power of this simplest framework, we derive predictions for the Dirac and Majorana CP phases in the lepton sector and for the observables which depend on the absolute mass scale of neutrinos. We also analyse the thermal leptogenesis within this set-up and outline its consequence on the derived predictions.

\section{CP invariance and Yukawa sum-rules}
\label{sec:CP}
The renormalizable Yukawa interactions of $16$-dimensional spinors $\psi_a$, containing the left chiral Weyl fermions, with a complex $\Phi$ and $\overline{\Sigma}$, which respectively transform as $10$ and $\overline{126}$ of $SO(10)$, can be written as \cite{Wilczek:1981iz}
\begin{widetext}
\be \label{LY}
-{\cal L}_Y =  (Y_{10})_{ab} \psi_a^T C^{-1} C_5 \Gamma_\mu \psi_b \Phi_\mu +  (\tilde{Y}_{10})_{ab} \psi_a^T C^{-1} C_5 \Gamma_\mu \psi_b \Phi^*_\mu + \frac{1}{5!} (Y_{\overline{126}})_{ab} \psi_a^T C^{-1} C_5 \Gamma_{[\mu} \Gamma_{\nu} \Gamma_{\rho} \Gamma_{\lambda} \Gamma_{\kappa]} \psi_b \overline{\Sigma}_{\mu\nu\rho\lambda\kappa}+{\rm h.c.},\ee
\end{widetext}
where $a,b = 1,2,3$ are flavour indices. $\Gamma_\mu$, with $\mu = 1,2,...,10$, are traceless matrices which define rank 10 Clifford algebra, $\{\Gamma_\mu,\Gamma_\nu\} = 2 \delta_{\mu \nu}  \mathbb{I}$. Also, $\Gamma_\mu^\dagger = \Gamma_\mu$ and $\Gamma_\mu^T = (-1)^{1+\mu} \Gamma_\mu$. $C$ is the usual charge conjugation matrix acting on Lorentz spinors and it obeys $C^T=C^\dagger=C^{-1} = -C$. Analogously, $C_5 = \Pi_{i = 1}^5 \Gamma_{2i - 1}$ is conjugation matrix which acts on $SO(10)$ spinors with properties: $C_5^\dagger = C_5^{-1} = C_5$ and $\Gamma_\mu C_5 = (-1)^{1+\mu}\,C_5 \Gamma_\mu$. It is well-known that the Fermi statistics along with the above properties of $\Gamma_\mu$ and $C_5$ imply that $Y_{10}$, $\tilde{Y}_{10}$ and $Y_{\overline{126}}$ are symmetric matrices \cite{Wilczek:1981iz} in flavour space.

We define the following CP transformations for the fermion and scalar fields.
\beqa \label{CP}
\psi_{a}(x) & \xrightarrow{\rm CP} & i  C \psi_{a}^*(\hat{x}), \nonumber \\
\Phi_\mu(x) & \xrightarrow{\rm CP} & (-1)^{1+\mu}\,\Phi^*_\mu(\hat{x}), \nonumber \\
\overline{\Sigma}_{\mu\nu\rho\lambda\kappa}(x) & \xrightarrow{\rm CP} & (-1)^{1+\mu} ... (-1)^{1+\kappa}\,\overline{\Sigma}_{\mu\nu\rho\lambda\kappa}^*(\hat{x}).\eeqa
where $x \equiv (x^0,x^i)$ and $\hat{x}\equiv (x^0,-x^i)$. The above transformations are specific choices from a more general class of CP transformations that leave the gauge interactions invariant \cite{Grimus:1995zi}. CP transforming each term in Eq. (\ref{LY}) using (\ref{CP}) and comparing the result with the hermitian conjugate of the original term, we find
\be \label{Y_CP}
Y_{ab} = Y^*_{ba},~~{\rm for}\, Y = Y_{10},\, \tilde{Y}_{10},\,Y_{\overline{126}}. \ee
The above along with their symmetric property implies that all the elements of $Y_{10}$, $\tilde{Y}_{10}$ and $Y_{\overline{126}}$ are real.

The scalar fields $\Phi$ and $\overline{\Sigma}$ each contains a pair of electroweak Higgs doublets, namely $h_1,\bar{h}_1 \in \Phi $ and $h_2,\bar{h}_2 \in \overline{\Sigma}$, where $h_{1,2}$ ($\bar{h}_{1,2}$) have hyperhcarge $Y=1/2$ ($-1/2$). The Yukawa interactions of the SM fermions and right-handed (RH) neutrinos with $h_{1,2}$ and $\bar{h}_{1,2}$, as computed from Eq. (\ref{LY}), are given by \cite{Mummidi:2021anm}
\beqa \label{LY2}
-{\cal L}_Y & \supset & \overline{Q}_L \left(2\sqrt{2} Y_{10}\,h_1 + 2\sqrt{2} \tilde{Y}_{10}\,\tilde{\bar{h}}_1 + 4\sqrt{\frac{2}{3}} Y_{\overline{126}}\,h_2 \right) d_R\, \nonumber \\
 &+& \overline{L}_L \left(2\sqrt{2} Y_{10}\,h_1 + 2\sqrt{2} \tilde{Y}_{10}\,\tilde{\bar{h}}_1 - 4\sqrt{6} Y_{\overline{126}}\,h_2 \right) e_R\, \nonumber \\
  &+& \overline{Q}_L \left(2\sqrt{2} Y_{10}\,\bar{h}_1 + 2\sqrt{2} \tilde{Y}_{10}\,\tilde{h}_1 + 4\sqrt{\frac{2}{3}} Y_{\overline{126}}\,\bar{h}_2 \right) u_R\, \nonumber \\
 &+& \overline{L}_L \left(2\sqrt{2} Y_{10}\,\bar{h}_1 + 2\sqrt{2} \tilde{Y}_{10}\,\tilde{h}_1 - 4\sqrt{6} Y_{\overline{126}}\,\bar{h}_2 \right) \nu_R\, \nonumber \\
 &+&  {\rm h.c.}, \eeqa
where $\tilde{h}_{1,2} = i\sigma_2 h_{1,2}^*$ and $\tilde{\bar{h}}_{1,2} = i\sigma_2 \bar{h}_{1,2}^*$. We have suppressed the flavour indices in writing the above. Note that the Higgs doublet residing in $\Phi$ couples to both the up-type and down-type fermions as both $\Phi$ and $\Phi^*$ participate in the Yukawa interaction. 

Next, we assume that only a linear combination of four Higgs doublets is light and participate in the electroweak symmetry breaking. In general, the Higgs doublets with identical SM charges mix with each other and their most general mass Lagrangian, in the basis ${\bf h} = \left(h_1,h_2,\tilde{\bar{h}}_1,\tilde{\bar{h}}_2\right)^T$, can be written as 
\be \label{LH}
{\cal L}^{\rm mass}_{\bf h} = -(M^2_{\bf h})_{ij}\, {\bf h}_i^\dagger {\bf h}_j\,.\ee
The diagonal elements of $M_{\bf h}^2$ can arise from the terms like $\Phi \Phi^*$ and $\overline{\Sigma} \overline{\Sigma}^*$ in the scalar potential and they are always real. The off-diagonal terms are in general complex parameters which can result from gauge invariant terms like $\Delta \Phi^2$, $\Delta^2 \Phi^* \overline{\Sigma}$, $\Delta^2 \Phi \overline{\Sigma}$ and/or $\Theta \Phi^* \overline{\Sigma}$, $\Theta \Phi \overline{\Sigma}$ when $\Delta$ and/or $\Theta$ acquires complex VEVs breaking the CP spontaneously. Here, $\Delta$ ($\Theta$) is 45-(210-)dimensional Lorentz scalar and at least one of them is necessarily required to be present in the complete model to break $SO(10)$ down to the SM \cite{Bertolini:2009qj}. A massless linear combination of ${\bf h}_i$ can be arranged through a usual fine-tuning condition, Det$(M_{\bf h}^2)=0$.

Identifying the lightest combination of ${\bf h}_i$ as $h$, the latter can be parametrized as 
\beqa \label{h}
h &=& c_\theta\, h_1 + s_\theta c_\chi e^{-i \phi_1}\, h_2 + s_\theta s_\chi c_\zeta e^{-i \phi_2}\, \tilde{\bar{h}}_1 \nonumber \\
&+& s_\theta s_\chi s_\zeta e^{-i \phi_3}\, \tilde{\bar{h}}_2\,,\eeqa
where $c_\theta = \cos\theta$, $s_\theta = \sin\theta$ and so on. Using Eqs. (\ref{LY2},\ref{h}), the Yukawa interactions of the SM quarks and leptons at the sub-GUT scale are then obtained as
\beqa \label{LY3}
-{\cal L}_Y & \supset & \overline{Q}_L\,Y_d\,h\,d_R\,+\, \overline{Q}_L\,Y_u\,\tilde{h}\,u_R\, \nonumber \\
&+& \overline{L}_L\,Y_{e}\,h\, e_R\,+\, \overline{L}_L\,Y_\nu\,\tilde{h}\,\nu_R\, +\, {\rm h.c.}\,, \eeqa
with
\beqa \label{Yf}
Y_d &=& H + r\, e^{i\phi_2} \tilde{H} + e^{i \phi_1} F, \nonumber \\
Y_e &=& H + r\, e^{i\phi_2} \tilde{H} -3\, e^{i \phi_1} F, \nonumber \\
Y_u &=& \tilde{H} + r\, e^{-i\phi_2} H + s\, e^{-i \phi_3} F, \nonumber \\
Y_\nu &=& \tilde{H} + r\, e^{-i\phi_2} H - 3s\, e^{-i \phi_3} F\,,\eeqa
and $H = 2\sqrt{2} c_\theta Y_{10}$, $\tilde{H} = 2\sqrt{2} c_\theta \tilde{Y}_{10}$, $F = 4 \sqrt{\frac{2}{3}} s_\theta c_\chi\, Y_{\overline{126}}$, $r=s_\chi c_\zeta\, \tan\theta$ and $s=s_\zeta\, \tan\chi$.

It is well-known that the third term in Eq. (\ref{LY}) can give masses to the RH neutrinos when an SM singlet and $B-L$ charged scalar, namely $\sigma$, residing in $\overline{\Sigma}$ acquires a VEV. We parametrise the resulting RH neutrino mass matrix as
\be \label{MR}
M_R = c\,v_{\sigma}\,Y_{\overline{126}} \equiv v_R\,F\,. \ee
The above, together with the Dirac neutrino masses $ v Y_\nu$, generates effectively the masses for the light neutrinos, through type I seesaw mechanism\footnote{There is also type II seesaw contribution induced through the VEV of electroweak triplet residing in $\overline{\Sigma}$. Depending on the choice of scalars and their potential in the full model, either of them can dominate over the other. We assume that type II contribution is negligibly small.}, given by
\be \label{Mnu}
M_\nu = -\frac{v^2}{v_R}\,Y_\nu\,F^{-1}\,Y_\nu^T\,,\ee
where $v \equiv \langle h \rangle \simeq 174$ GeV.

The Yukawa sum rules derived in Eq. (\ref{Yf}) and $M_\nu$ given in Eq. (\ref{Mnu}) determine all the masses and mixing observables of the SM fermions including the CP violation in both the quark and lepton sectors. The model contains 21 real parameters (see the next section for the count)  in comparison to 21 in \cite{Grimus:2006bb} and 17 in \cite{Grimus:2006rk,Joshipura:2011nn}  proposed and analysed earlier based on $SO(10)$ and spontaneous CP violation. However, all the previous frameworks use the extended Yukawa sector including $120$-dimensional Higgs and additional symmetries like $Z_2$ or $U(1)$, or assume purely imaginary VEVs. The model proposed here does not rely on such ad-hoc assumptions.

The most noteworthy feature of the relations, Eqs. (\ref{Yf},\ref{Mnu}), is that the CP violation arises only from phases $\phi_k$ ($k=1,2,3$) since all the remaining parameters are real. The model, therefore, provides a  predictive setup in which the CP violation in the quark and lepton sectors is expected to be correlated.

\section{Fermion spectrum fit and predictions}
\label{sec:fit}
With a suitable choice for the basis of $\psi_a$ in Eq. (\ref{LY}), the matrix $\tilde{H}$ in Eq. (\ref{Yf}) can be made diagonal and positive.  This leaves a total of 21 real parameters (3 in $\tilde{H}$, 6 each in $H$ and $F$, three phases $\phi_k$ and $r$, $s$, $v_R$) which determine 22 observable quantities that include 12 fermion masses, 6 mixing angles in the quark and lepton sectors and 4 CP phases. Out of these, the absolute scale of the neutrino masses and three leptonic CP phases are not yet directly measured\footnote{Indirect limits exist on values of the Dirac CP phase $\delta$ through the global fits to neutrino oscillation data \cite{deSalas:2020pgw,Esteban:2020cvm,Capozzi:2017ipn}. However, they allow an almost entire range of $\sin\delta$ at $3\sigma$.}. Our aim in this section is to fit the 18 experimentally observed quantities to determine the 21 unknown parameters using the $\chi^2$-minimization method \cite{Mummidi:2021anm} and use the obtained parameters to derive predictions for the unmeasured observables.

Our numerical analysis method is similar to the one used earlier and described at length in \cite{Mummidi:2021anm}. The fit is carried out for the values of parameters at the GUT scale, $M_{\rm GUT} = 2 \times 10^{16}$ GeV. We use 1-loop renormalization group evolution (RGE) equations of the SM to determine the charged fermion masses and quark mixing parameters at $M_{\rm GUT}$. For the neutrino masses, we assume normal ordering which is qualitatively plausible considering the quark-lepton unification. It is known that RGE effects in the neutrino masses and leptonic mixing angles are small for the normal ordering and, therefore, we use low-energy values for these observables from \cite{Esteban:2020cvm} for fitting. The values of various observables used in $\chi^2$ are given as $O_i^{\rm exp}$ in Table \ref{tab:bestfit}. For the standard deviation, we consider $30 \%$ for the light quark masses and $10 \%$ for all the remaining observables. These conservative uncertainties are taken to account for higher-order RGE effects and threshold corrections. We also assume no sizeable modification in the running of Yukawa couplings due to possible new intermediate scales. Actual incorporation of such effects requires the specification of a complete Higgs sector and symmetry-breaking pattern beyond the Yukawa sector and is highly model-dependent. Nevertheless, with the use of conservative standard deviations, we expect our results to not get drastically altered in the presence of such effects.

\begin{figure}[t]
\centering
\subfigure{\includegraphics[width=0.32\textwidth]{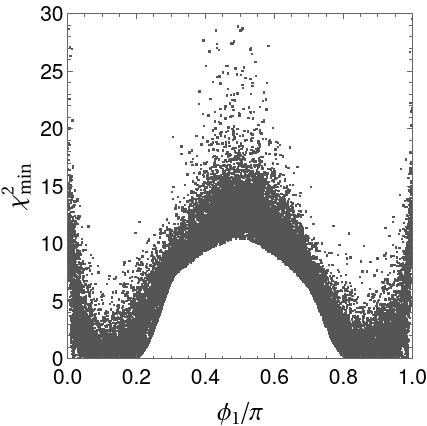}}
\caption{Minimized $\chi^2$ for different values of the phase $\phi_1$.}
\label{fig1}
\end{figure}
Since $\phi_k$ are the only source of non-vanishing CP, we first randomly vary their values (which can be taken between $0$ and $\pi$ without loss of generality) simultaneously and optimize the values of the other 18 parameters using $\chi^2$ minimization. The result is displayed in Fig. \ref{fig1} in case of $\phi_1$ for which we get a clear correlation. Solutions with $\chi^2_{\rm min} < 9$  disfavours $0.4\pi < \phi_1 < 0.6\pi$. In case of $\phi_{2,3}$, we do not find any specific correlation. Almost all the values of $\phi_{2,3}$ are found to give acceptable $\chi^2_{\rm min}$. We also find $\chi^2_{\rm min} \geq 100$ for $\phi_k = 0$ or $\pi$, since they cannot reproduce the non-zero CKM phase. Although, even a very small deviation from these values leads to a good fit.  Out of 40K distinct solutions obtained for the full range of $\phi$ shown in Fig. \ref{fig1},  approximately 23K are found with $\chi_{\rm min}^2 \le 9$. This choice of  $\chi_{\rm min}^2$ ensures that no observable deviates more than $3\sigma$ from its central value and hence they can be considered viable solutions. The reproduced values of the observables and optimized values of the parameters for one of the best-fit solutions, corresponding to $\chi^2_{\rm min} = 0.11$, are given in Appendix \ref{app:bestfit} for illustration.

\begin{figure*}[t]
\centering
\subfigure{\includegraphics[width=0.29\textwidth]{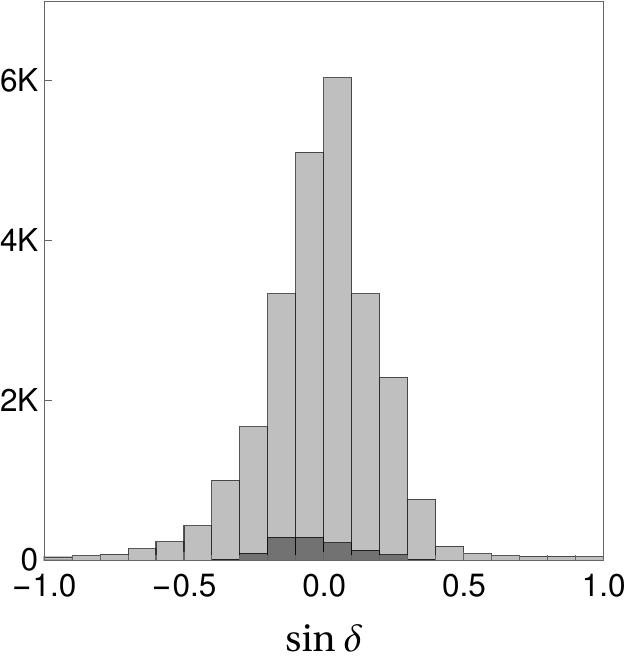}}\hspace*{0.1cm}
\subfigure{\includegraphics[width=0.29\textwidth]{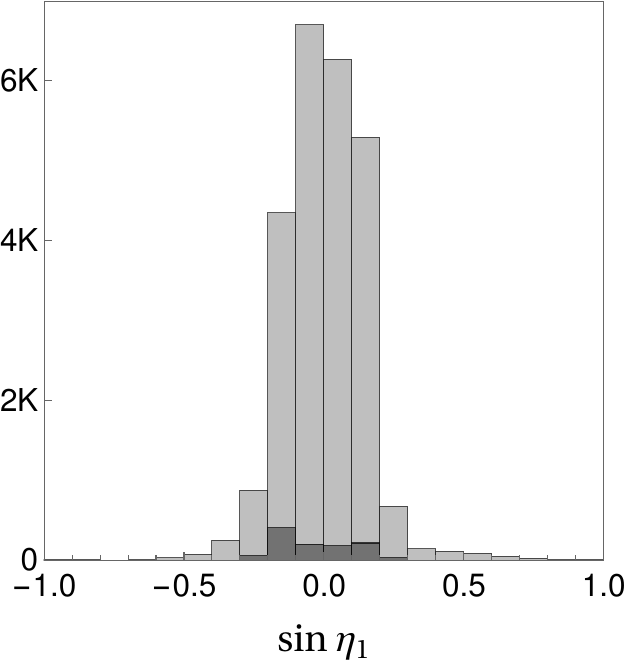}}\hspace*{0.1cm}
\subfigure{\includegraphics[width=0.29\textwidth]{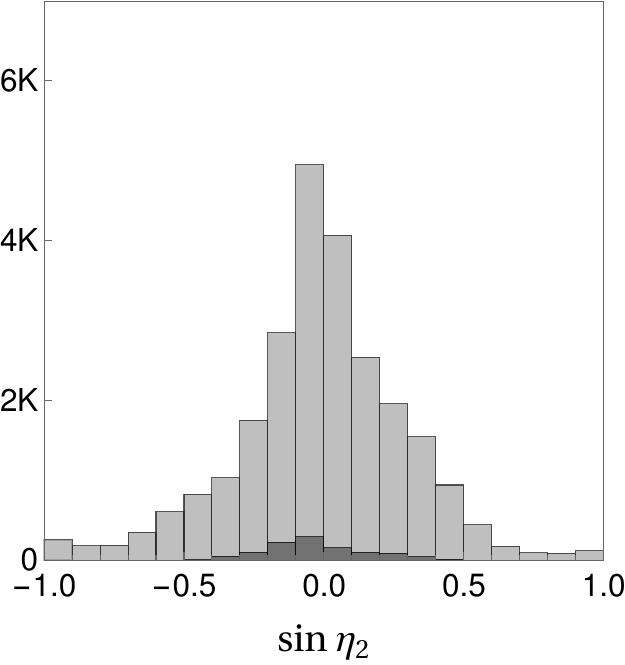}}
\subfigure{\includegraphics[width=0.28\textwidth]{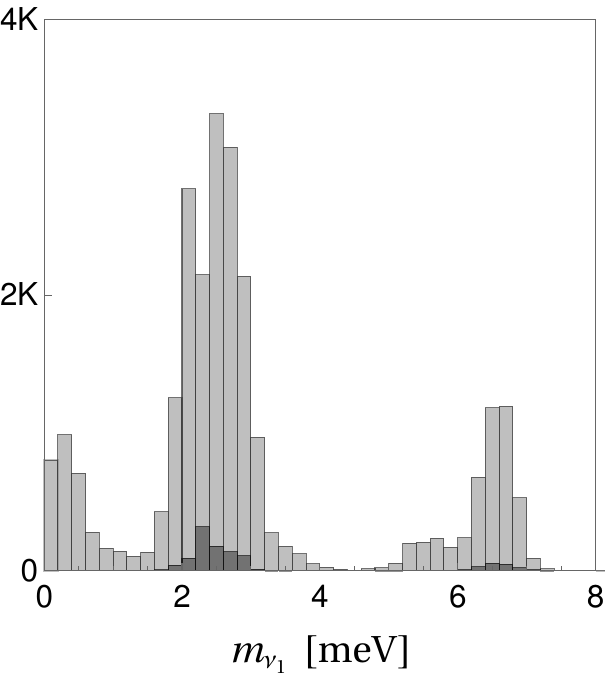}}\hspace*{0.2cm}
\subfigure{\includegraphics[width=0.285\textwidth]{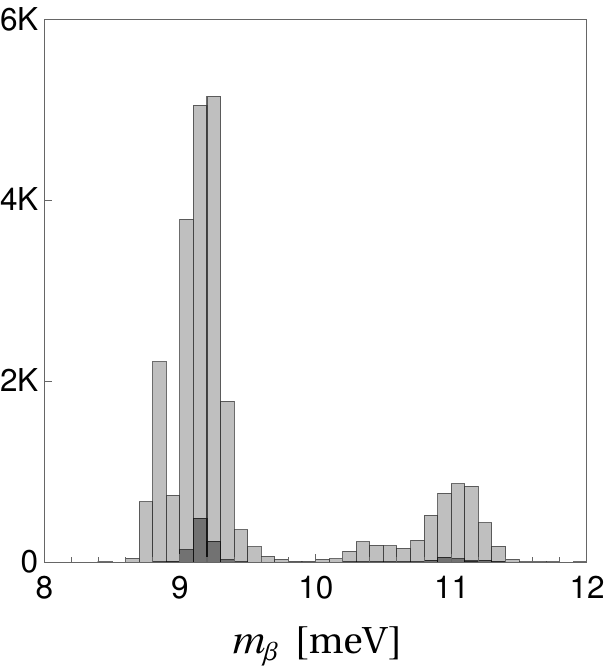}}\hspace*{0.2cm}
\subfigure{\includegraphics[width=0.295\textwidth]{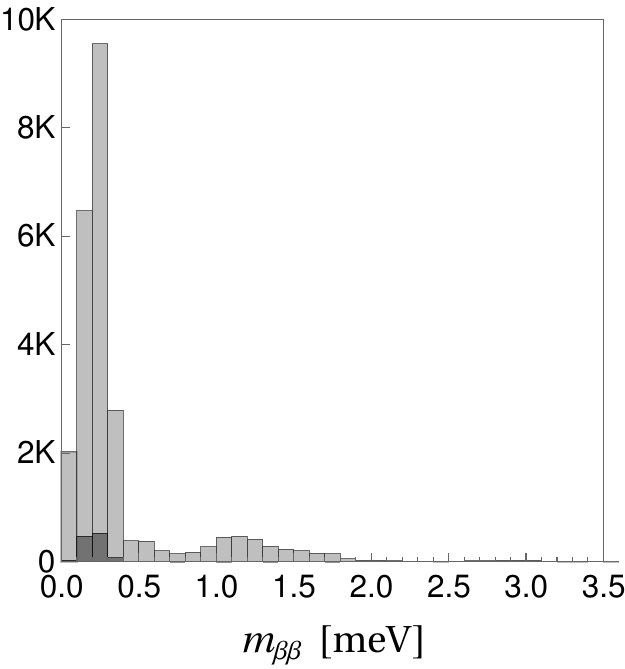}}
\caption{Number of solutions with $\chi^2_{\rm min} \le 9$ as a function of the Dirac CP phase ($\delta$), Majorana phases ($\eta_{1,2}$),  the lightest neutrino mass ($m_{\nu_1}$), the effective beta decay mass ($m_\beta$) and the effective Majorana mass of the electron-neutrino ($m_{\beta \beta}$). The darker bars in all the plots correspond to the solutions which also reproduce viable baryon asymmetry through leptogenesis.}
\label{fig2}
\end{figure*}
To derive comprehensive predictions, we do not rely only on the best fit solution and consider all the solutions with  $\chi_{\rm min}^2 \le 9$. We compute the leptonic Dirac phase, $\delta$, using the Jarlskog invariant \cite{Jarlskog:1985ht} and the standard parametrization of the lepton mixing matrix $U$ given in the PDG \cite{ParticleDataGroup:2022pth}. For the Majorana CP phases, we use the rephasing invariants \cite{Branco:1986gr,Nieves:1987pp,Jenkins:2007ip}
\beqa \label{I12}
{\cal I}_1 & = & {\rm Im}\left[ U_{11}^2 U_{13}^{*2} \right] = c_{12}^2 c_{13}^2 s_{13}^2 \sin 2(\eta_1 + \delta), \nonumber \\
{\cal I}_2 & = & {\rm Im}\left[ U_{12}^2 U_{13}^{*2} \right] = s_{12}^2 c_{13}^2 s_{13}^2 \sin 2(\eta_2 + \delta), \eeqa
where the second equality is obtained using again the standard parametrization of the unitary matrix $U$.  We also compute the other relevant predictions which include the mass of the lightest neutrino $m_{\nu_1}$, the effective mass of the electron-neutrino $m_\beta = \sqrt{\sum_i m_{\nu_i}^2 |U_{1i}|^2}$ and the effective Majorana mass of electron-neutrino $m_{\beta \beta} = \left|\sum_i m_{\nu_i} U_{1i}^2 \right|$, respectively. $m_\beta$ and $m_{\beta \beta}$ can be measured directly in the beta decay and neutrinoless double beta decay experiments. The results are shown in Fig. \ref{fig2} which is self-explanatory. Unlike the previous scenarios \cite{Joshipura:2011nn,Feruglio:2014jla,Babu:2018tfi,Mummidi:2021anm}, the present framework shows clear preferences for certain ranges of the Dirac and Majorana CP phases.

\section{Leptogenesis}
\label{sec:lepto}
\begin{figure}[t]
\centering
\subfigure{\includegraphics[width=0.29\textwidth]{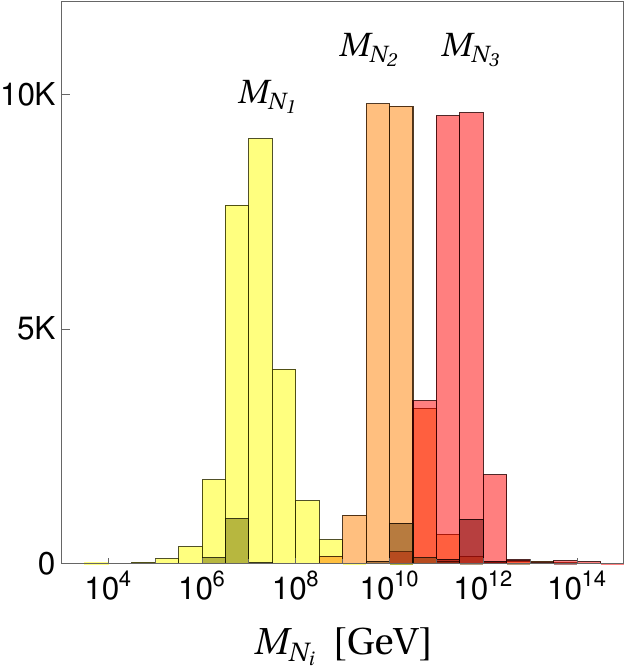}}
\caption{The mass spectrum of RH neutrinos as predicted by the model. The small darker bars correspond to the solutions which also reproduce viable baryon asymmetry through leptogenesis.}
\label{fig3}
\end{figure}

The CP-violating out-of-equilibrium decays of the RH neutrinos can give rise to lepton asymmetry \cite{Fukugita:1986hr} which subsequently can get converted into the baryon asymmetry of the universe through the electroweak sphaleron processes, see \cite{Bodeker:2020ghk,DiBari:2021fhs} for recent reviews. The precise computation of lepton asymmetry depends on the masses of RH neutrinos. We, therefore, compute the mass spectrum of the RH neutrinos in the model for all the viable solutions. The results are displayed in Fig. \ref{fig3}. 

Since $M_{N_1} \ll M_{N_2} < 10^{12}$ GeV, the lepton and anti-lepton flavour states which couple to the RH neutrinos do not maintain the coherence between their production and inverse decays. Moreover, in this case, one must also consider the contributions to the final lepton asymmetry from the RH neutrinos heavier than $N_1$ \cite{Vives:2005ra,DiBari:2005st,Abada:2006ea,Nardi:2006fx,Engelhard:2006yg}. Therefore, it becomes necessary to compute kinetic evolutions of the lepton asymmetry using the Density Matrix Equations (DME) instead of the classical Boltzmann equations.

The most general DME applicable for three RH neutrinos and valid for $M_{N_1} > 10^6$ GeV are derived in \cite{Blanchet:2011xq} which we use for the evaluation of the total $B-L$ asymmetry, namely $N^{\rm f}_{B-L}$. The latter is given by a trace of asymmetry matrix $N_{\alpha \beta}$ where $\alpha$ and $\beta$ denote lepton flavours. $N_{\alpha \beta}$ can be computed by solving the DME numerically. For this analysis, we follow the procedure and method described in detail in \cite{Mummidi:2021anm}.  The final baryon-to-photon ratio is computed using 
\be \label{etaB}
\eta_B = 0.0096\, N^{\rm f}_{B-L}\,,\ee
where the numerical factor takes into account the sphaleron conversion and dilution due to an increase in the number of photons in the comoving volume.

Although we compute $N^{\rm f}_{B-L}$ numerically, an approximate analytical solution of the same, valid for the RH neutrino mass spectrum predicted by the model, is given by
\beqa \label{NBL_ana}
N^{\rm f}_{B-L} & \simeq & \varepsilon_{1 \tau}\, \kappa(K_{1 \tau}) + \varepsilon_{1 \tau^\perp}\, \kappa(K_{1 \tau^\perp}) \nonumber \\
&+& p_{12}\, \varepsilon_{2 \tau^\perp}\, \kappa(K_{2 \tau^\perp})\, e^{-\frac{3 \pi}{8}K_{1 \tau^\perp}}\nonumber \\
&+& (1-p_{12})\, \varepsilon_{2 \tau^\perp}\, \kappa(K_{2 \tau^\perp}).\eeqa
The various quantities appearing above are defined in \cite{Mummidi:2021anm}. The above is derived systematically following the steps discussed in \cite{Mummidi:2021anm}. The first two terms in Eq. (\ref{NBL_ana}) denote asymmetry produced by decays of $N_1$ in the $\tau$-lepton flavour state and its orthogonal state, respectively.  The term in the second line is a fraction of asymmetry produced in $N_2$ decays that gets diluted due to washout processes involving $N_1$. The last term is the fraction of asymmetry which does not get affected by $N_1$ washouts due to the flavour effects.

We numerically solve the full set of DME and find that the mismatch between the analytical and exact numerical results is less than ${\cal O}(100 \%)$ for almost all the points with $\chi^2_{\rm min} \le 9$. Therefore, the analytical solution may not be suitable for the accurate determination of $\eta_B$. Nevertheless, an extremely simple form of Eq. (\ref{NBL_ana})  allows one to make a quick estimation for an order of magnitude of $\eta_B$ in this class of models.

Out of the $23$K solutions with viable fit to the fermion mass spectrum, we find that around $1.1$K solutions reproduce values of $\eta_B$ in the range $6.12 \times 10^{-9}$ - $6.12 \times 10^{-11}$. We have chosen a conservative range instead of the experimentally measured value $\eta_B^{\rm exp} = (6.12 \pm 0.04) \times 10^{-10}$ since the Yukawa couplings and RH neutrino spectrum which determine $\eta_B$ are obtained with $10\%$-$30\%$ uncertainties in the other observables. These solutions are indicated in Figs. \ref{fig2} and \ref{fig3} by darker bars. It can be seen that the requirements of viable baryon asymmetry through the thermal leptogenesis within the underlying model significantly narrow down the range of predicted quantities. The results show that successful leptogenesis can be realized in the model despite of relatively small CP violation in the lepton sector.

\section{Conclusion}
\label{sec:concl}
We propose a minimal renormalizable Yukawa sector for non-supersymmetric $SO(10)$ GUT which uses CP symmetry. The latter is broken spontaneously leading to the effective quark and lepton Yukawa matrices that contain only three phases. Observed CP violation in the quark sector restricts the range of one of these phases, namely $\phi_1$, which in turn leads to predictions for CP phases in the lepton sector. From comprehensive fits to the fermion masses and mixing parameters, we find that the framework shows preference for relatively small CP violation in the lepton sector. The model also has specific predictions for the lightest neutrino mass ($m_{\nu_1} \in [0,0.007]$ eV), the rates of double beta ($m_{\beta} \in [0.0085,0.0115]$ eV) and neutrinoless double beta decay ($m_{\beta \beta} \in [10^{-4},0.002]$ eV). The requirement of successful leptogenesis within this model further reduces the ranges of predicted observables. In particular, we find $\sin\delta \in [-0.4,0.4]$ for the Dirac and $\sin\eta_1 \in [-0.3,0.2]$ and $\sin\eta_2 \in [-0.5, 0.5]$ for the Majorana CP phases. The conclusive predictions can be tested in the neutrino oscillation and non-oscillation experiments making the underlying model a falsifiable framework of gauge and quark-lepton unification.

\section*{Acknowledgements}
We are grateful to an anonymous reviewer for pointing out an important error in the previous version. This work is partially supported under the MATRICS project (MTR/2021/000049) funded by the Science \& Engineering Research Board (SERB), Department of Science and Technology (DST), Government of India.

\bibliography{references}
\bibliographystyle{apsrev4-1}

\pagebreak
\onecolumngrid

\appendix
\section{Best fit solution}
\label{app:bestfit}
In this Appendix, we give details of one of the best-fit solutions corresponding to $\chi^2_{\rm min} = 0.11$. There are several solutions with lesser $\chi^2_{\rm min}$, however, the present is the best fit solution which also reproduces the viable value of $\eta_B$. The optimized values of the parameters are obtained as:
\be \label{tildeH_sol}
\tilde{H} = \left(
\begin{array}{ccc}
 - 2.9464 \times 10^{-6} & 0 & 0 \\
 0 & 0.00146793 & 0 \\
 0 & 0 & 0.444117 \\
\end{array}
\right),\, 
H = \left(
\begin{array}{ccc}
 0.000120191 & 0.027934 & 0.000566981 \\
 0.027934 & -0.107859 & 0.457788 \\
 0.000566981 & 0.457788 & -1.25006 \\
\end{array}
\right) \times 10^{-3}, \ee
\be \label{F_sol}
F = \left(
\begin{array}{ccc}
 3.5123\times 10^{-7} & 1.26879 \times 10^{-4} & 0.0250325 \\
1.26879 \times 10^{-4} & 0.0144722 & -0.545796 \\
 0.0250325 & -0.545796 & -3.50025 \\
\end{array}
\right) \times 10^{-3}, \ee
\be \label{ph_sol}
\phi_1 = 0.554938,\,\phi_2 =2.38024 ,\, \phi_3 = 1.24964\,, \ee
\be \label{ph_rs}
r= 4.98311 \times 10^{-3},\,s= -9.98489 \times 10^{-4}, \,v_R = 1.94476 \times 10^{14}\, {\rm GeV}.\ee

All the observables of quark and lepton mass spectrum can be determined by substituting the above values in Eqs. (\ref{Yf},\ref{Mnu}). $\eta_B$ can be computed by solving the DME following the procedure explicitly outlined in \cite{Mummidi:2021anm}. The computed values of various observables are given as $O_i^{\rm th}$ in Table \ref{tab:bestfit}. $O_i^{\rm exp}$ is the reference value used in the fit and their details are given in the main text. The pull, $(O_i^{\rm th}-O_i^{\rm exp})/\sigma_i$, denotes deviation in the fitted observables. Predictions corresponding to this solution are also listed in Table \ref{tab:bestfit}.
\begin{table}[h]
	\begin{center} 
		\begin{math} 
			\begin{tabular}{cccc}
				\hline
				\hline 
				~~~Observable~~~ & ~~~~~~~~$O_i^{\rm th}$~~~~~~~~  & ~~~~~~~~$O_i^{\rm exp}$~~~~~~~~  & ~~~Pull~~~ \\
				\hline
				$y_u$  & $2.95 \times 10^{-6}$ & $2.92\times 10^{-6}$ & $\sim 0$ \\
				$y_c$   & $1.47\times10^{-3}$& $1.47\times 10^{-3}$& $0$\\
				$y_t$   & $0.444 $ & $0. 444$& $0$\\
				$y_d$  & $ 6.87 \times 10^{-6} $ & $6.42\times10^{-6}$  & $0.2$ \\
				$y_s$   & $1.22 \times10^{-4}$ & $1.28\times10^{-4}$ & $-0.2$ \\
				$y_b$   & $5.85 \times10^{-3}$ & $5.86 \times10^{-3}$& $\sim 0$ \\
				$y_e$   & $2.75 \times 10^{-6}$ & $2.76 \times 10^{-6}$ & $\sim 0$ \\
				$y_{\mu}$  &$5.75 \times10^{-4}$ & $5.75\times10^{-4}$ & $0$\\
				$y_{\tau}$   & $9.73 \times10^{-3}$ & $9.72 \times10^{-3}$& $\sim 0$\\
				$\Delta m^2_{\text{sol}} [{\rm eV}^2]$ & $7.40\times10^{-5}$& $7.42\times10^{-5}$ & $\sim 0$\\
				$\Delta m^2_{\text{atm}} [{\rm eV}^2]$ & $2.515\times10^{-3}$& $2.515\times10^{-3}$& $0$\\
				$|V_{us}|$ & 0.2273 & 0.2304 & $-0.1$\\
				$|V_{cb}|$ & 0.0483 & 0.0483 & $0$ \\
				$|V_{ub}|$ & 0.0043 & 0.0043 & $0$ \\
				$\sin\delta_{\rm CKM}$ & 0.909 & 0.910 & $\sim 0$  \\
				$\sin^2 \theta _{12}$ & 0.302 & 0.304 & $-0.1$\\
				$\sin^2 \theta _{23}$ & 0.576   & 0.573 & $0.1$ \\
				$\sin^2 \theta _{13}$ & 0.02229 & 0.02220  & $\sim 0$ \\	  	
				\hline
				\multicolumn{4}{c}{Predictions} \\
				\hline
				$\sin\delta$ & -0.205 & $M_{N_1}$ [GeV]  & $7.02 \times 10^{6}$ \\
				$\sin\eta_1$ & -0.183  & $M_{N_2}$ [GeV]  & $1.89 \times 10^{10}$ \\
				$\sin\eta_2$ & -0.205 & $M_{N_3}$ [GeV]  & $ 6.97 \times 10^{11}$ \\
				$m_{\nu _1}$[meV] & 2.46  & $\eta_B$ & $5.96 \times 10^{-10}$\\
				$ m_{\beta }$[meV]  & 9.17 & & \\		
				$m_{\beta \beta}$[meV]  & 0.17 & & \\			
				\hline
				\hline 
			\end{tabular}
		\end{math}
	\end{center}
	\caption{Results and predictions obtained for an example solution corresponding to $\chi_{\rm min}^2 = 0.11$.} 
	\label{tab:bestfit} 
\end{table}

\end{document}